\documentclass[aps,prl,10pt,twocolumn,floatfix,superscriptaddress,preprintnumbers,shownopacs,showkeys,nofootinbib,notoccite,notitlepage]{revtex4-2}
\usepackage{slashed}
\usepackage{graphicx}
\usepackage{dcolumn}
\usepackage{footmisc}
\usepackage{bm}
\usepackage[colorlinks = true, linkcolor = purple, urlcolor  = cyan, citecolor = blue, anchorcolor = blue]{hyperref}
\usepackage{xcolor}
\usepackage{float}
\usepackage{amsmath}

\def\DM{\mathrm{DM}}
\def\SM{\mathrm{SM}}
\usepackage{tikz,xcolor,hyperref}
\definecolor{darkgreen}{rgb}{0.0, 0.2, 0.13}
\definecolor{bostonuniversityred}{rgb}{0.8, 0.0, 0.0}
\definecolor{lime}{HTML}{A6CE39}
\DeclareRobustCommand{\orcidicon}{
	\begin{tikzpicture}
	\draw[lime, fill=lime] (0,0) 
	circle [radius=0.16] 
	node[white] {{\fontfamily{qag}\selectfont \tiny ID}};
	\draw[white, fill=white] (-0.0625,0.095) 
	circle [radius=0.007];
	\end{tikzpicture}
	\hspace{-2mm}
}
\foreach \x in {A, ..., Z}{\expandafter\xdef\csname orcid\x\endcsname{\noexpand\href{https://orcid.org/\csname orcidauthor\x\endcsname}
			{\noexpand\orcidicon}}
}
\begin{document}

\title{Cosmic-Ray Cooling in Active Galactic Nuclei as a New Probe of Inelastic Dark Matter}
\author{R. Andrew Gustafson\hspace{-1mm}\orcidA{}}
\email{gustafr@vt.edu}
\affiliation{Center for Neutrino Physics, Department of Physics, Virginia Tech, Blacksburg, Virginia 24061, USA}
\affiliation{International Center for Quantum-field Measurement Systems for Studies of the Universe and Particles (QUP,WPI), High Energy Accelerator Research Organization (KEK), Oho 1-1, Tsukuba, Ibaraki 305-081, Japan}

\author{Gonzalo Herrera\hspace{-1mm}\orcidB{}}
\email{gonzaloherrera@vt.edu}
\affiliation{Center for Neutrino Physics, Department of Physics, Virginia Tech, Blacksburg, Virginia 24061, USA}

\author{Mainak Mukhopadhyay\hspace{-1mm}\orcidC{}}
\email{mkm7190@psu.edu}
\affiliation{Department of Physics; Department of Astronomy \& Astrophysics; Center for Multimessenger Astrophysics, Institute for Gravitation and the Cosmos, The Pennsylvania State University, University Park, Pennsylvania 16802, USA}

\author{Kohta Murase\hspace{-1mm}\orcidD{}}
\email{murase@psu.edu}
\affiliation{Department of Physics; Department of Astronomy \& Astrophysics; Center for Multimessenger Astrophysics, Institute for Gravitation and the Cosmos, The Pennsylvania State University, University Park, Pennsylvania 16802, USA}
\affiliation{Center for Gravitational Physics and Quantum Information, Yukawa Institute for Theoretical Physics, Kyoto, Kyoto 606-8502 Japan}

\author{Ian M. Shoemaker\hspace{-1mm}\orcidE{}}
\email{shoemaker@vt.edu}
\affiliation{Center for Neutrino Physics, Department of Physics, Virginia Tech, Blacksburg, Virginia 24061, USA}

\begin{abstract}
We present a novel way to probe inelastic dark matter using cosmic-ray (CR) cooling in active galactic nuclei (AGNs). Dark matter (DM) in the vicinity of supermassive black holes may scatter off CRs, resulting in the rapid cooling of CRs for sufficiently large cross sections. This in turn can alter the high-energy neutrino and gamma-ray fluxes detected from these sources. We show that AGN cooling bounds obtained through the multimessenger data of NGC 1068 and TXS 0506+056 allows us to reach unprecedently large mass splittings for inelastic DM ($\gtrsim$ TeV), orders of magnitude larger than those probed by direct detection experiments and DM capture in neutron stars. Furthermore, we demonstrate that cooling bounds from AGNs can probe thermal light DM with small mass splittings. This provides novel and complementary constraints in parts of a parameter space accessible solely by colliders and beam dump experiments.

\end{abstract}

\maketitle

\emph{Introduction.-}
A pressing problem in high-energy physics and cosmology resides in the yet unknown nature of dark matter (DM), confirmed only via its gravitational effects on visible matter \cite{Green:2021jrr}. In the current paradigm, the DM is believed to likely be composed of one or more fundamental particles, that couple weakly or feebly to the Standard Model (SM) sector \cite{Bertone:2004pz,Profumo:2019ujg,Cirelli:2024ssz}. 

An early proposal to search for weakly interacting massive particles accounting for the observed DM abundance of the Universe, dubbed direct detection, consists in looking for its scatterings off nuclei at Earth-based detectors \cite{Goodman:1984dc, Drukier:1986tm}. In some DM models, the inelastic scattering channel can naturally dominate over the elastic one \cite{Hall:1997ah,Tucker-Smith:2001myb,Arkani-Hamed:2008hhe,Alves:2009nf,Chang:2010en,Nagata:2014wma,Nagata:2014aoa,Filimonova:2022pkj, Brahma:2023psr}. In this scenario, DM with mass $m_{\rm DM}$ upscatters with SM particles to an excited (heavier) state with mass $m_{\rm DM}^*$, where $m_{\rm DM}^* = m_{\rm DM} + \delta_{\rm DM}$ and $\delta_{\rm DM}$ is defined as the mass splitting.
A canonical example is the vector current of Majorana DM, which is forbidden for the elastic case, but not for the inelastic one. Indeed, for a Majorana fermion $\psi$,
\begin{equation}
\overline{\psi} \gamma_\mu \psi=\overline{\psi^c} \gamma_\mu\psi^c=-\overline{\psi} \gamma_\mu \psi,
\end{equation}
since for a Majorana field the charge conjugation operation leaves the field unchanged $\psi^c=\psi$. The off-diagonal current between two non-degenerate Majorana fields, could, however, be nonzero (see the Supplementary Material for a discussion on concrete realizations of inelastic DM. \cite{Tucker-Smith:2001myb,Garcia:2024uwf,Kahlhoefer:2015bea,Duerr:2016tmh}).

Such inelastic DM models are only weakly constrained by direct detection experiments, reaching maximum mass splittings between the two DM states of order $\sim$ 100 keV, e.g., Refs.~\cite{Tucker-Smith:2001myb, Schwetz:2011xm,Bramante:2016rdh, Baryakhtar:2020rwy, Song:2021yar, Bell:2021xff, Herrera:2023fpq, Eby:2023wem, Chatterjee:2022gbo, Kang:2024kec}, and are largely unconstrained by direct detection for sub-GeV DM masses, e.g., Refs.~\cite{Emken:2021vmf, Herrera:2023fpq, Garcia:2024uwf, Essig:2022dfa, Yun:2023huf, Li:2022acp, Bell:2022dbf, Gu:2022vgb}.
\begin{figure}[ht!]
    \centering
	\includegraphics[width=0.49\textwidth]{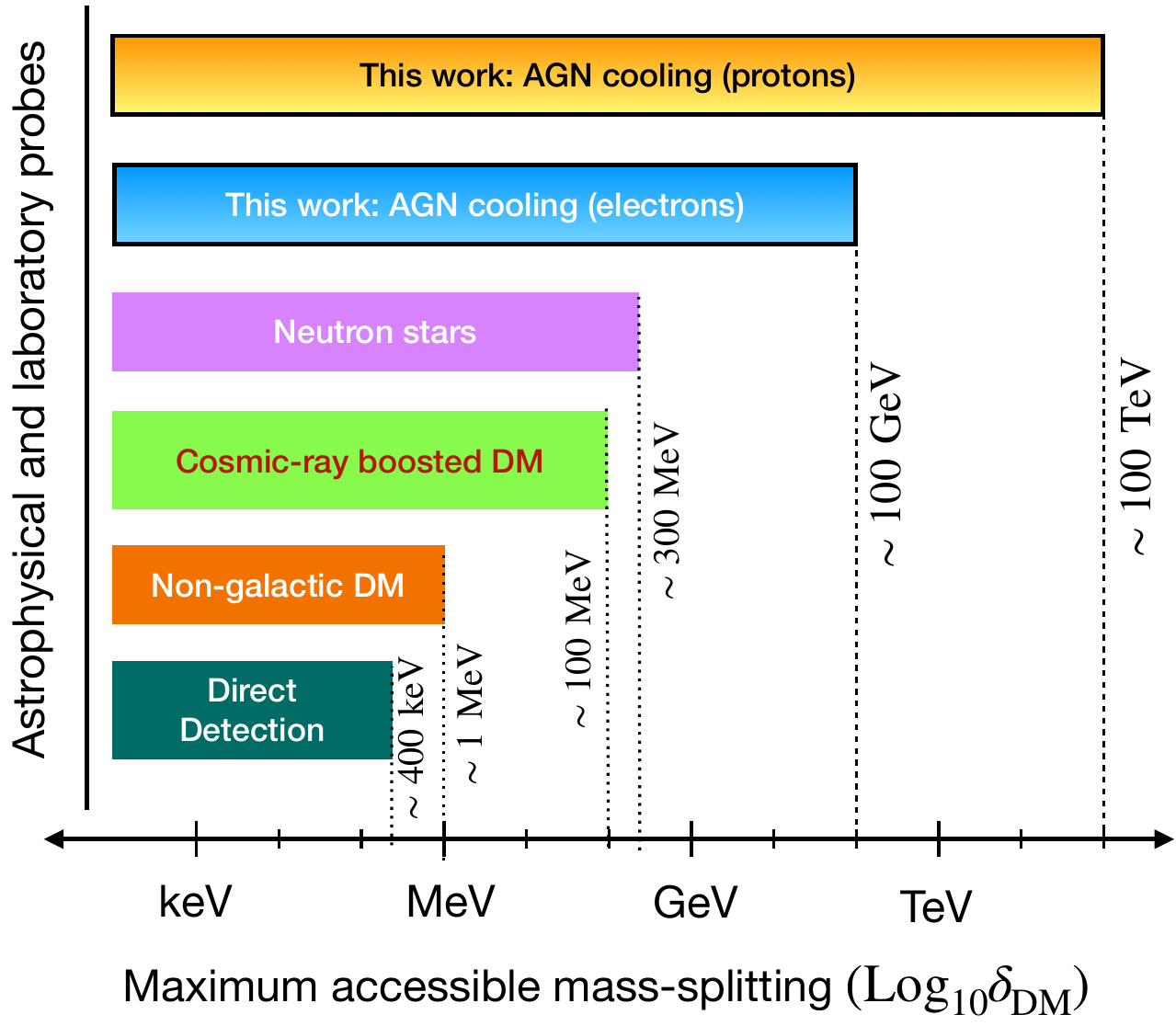}
	\centering
	\caption{Maximum mass splitting of inelastic DM reached by various astrophysical and laboratory probes: direct detection of DM from the galactic halo \cite{Song:2021yar}, direct detection of a nongalactic high-speed DM component \cite{Herrera:2023fpq}, direct detection of cosmic-ray (CR) boosted DM \cite{Bell:2021xff}, DM capture in neutron stars \cite{Bell:2018pkk}, and CR cooling in AGN (this work).  
    The cooling of CRs in AGN, inferred from multimessenger high-energy neutrino and electromagnetic observations, allows us to reach the largest mass splittings of inelastic DM to date.}
	\label{fig:schematic}
\end{figure}
Furthermore, it should be noted that indirect (astrophysical) constraints on inelastic DM restrict to some regions of parameter space only (and rely on future observations of nearby neutron stars), e.g., Refs.~\cite{McCullough:2010ai,Hooper:2010es, Bell:2018pkk, Alvarez:2023fjj, Chauhan:2023zuf, Biswas:2022cyh, Fujiwara:2022uiq, Acevedo:2024ttq}, or probe self-annihilations of DM particles, but lack in probing signatures arising from scatterings~\cite{Berlin:2023qco}.

In some regions of parameter space, focused particularly on comparable masses of the DM and the particle mediating the interaction between the DM and the SM (``mediator'' hereafter), colliders and fixed-target experiments can produce strong constraints on a dark sector mediator, e.g., Refs.~\cite{Berlin:2018bsc, Mongillo:2023hbs, Garcia:2024uwf}. While collider constraints provide complementary constraints to direct detection and astrophysical probes, they cannot probe particles with cosmological lifetimes nor the distribution of DM in the Universe. Besides, previous analyses have not been assessed for sub-MeV DM masses, nor have studied large ratios between the DM mass and the mass splitting.

Here we propose the cooling of cosmic rays (CRs) from some active galactic nuclei (AGN), inferred from combined electromagnetic (EM) and high-energy neutrino observations, as a probe of inelastic DM across orders of magnitudes in DM mass, mediator mass and mass splitting. Ref.~\cite{Herrera:2023nww} demonstrated that light DM in the vicinity of the black hole scattering off CR protons and electrons can cool them in such environments, affecting their multimessenger emissions. The impact of light DM-CR scatterings on only the electromagnetic emission from the nuclei of starbust galaxies was also considered in Ref. \cite{Ambrosone:2022mvk}. We will demonstrate here that DM-proton and DM-electron upscattering the DM to an excited state can also allow us to constrain inelastic DM, probing new regions of parameter space, and filling gaps in current constraints obtained with complementary probes such as direct detection and collider experiments.

\emph{CR cooling timescales in AGN.-}
CR protons and electrons can be efficiently accelerated in the vicinity of a central supermassive black hole through shocks, turbulence or magnetic reconnections. Plausible acceleration sites include disk-coronae and jets~\cite{Rieger:2022qhs}. The accelerated CR protons then interact with the protons or the photons in the respective regions to produce neutrinos and EM signatures through $pp$ and $p\gamma$ processes~\cite{Murase:2022feu}. The CR protons also cool through various other SM processes like synchrotron, inverse Compton, Bethe-Heitler pair production processes and adiabatic losses. Furthermore, the escape of CR protons or electrons from the sources are quantified by advection or diffusion. For NGC 1068 and TXS 0506+056, detailed multimessenger data and modeling are available, enabling us to evaluate the energy dependence of cooling times. This allows us to constrain DM properties only via bolometric luminosities without relying on intrinsic spectra of astrophysical neutrinos that are currently uncertain.  

CR protons in NGC 1068 are cooled at high energies mainly via $pp$, $p\gamma$ and Bethe-Heitler interactions ($N\gamma \rightarrow N^{*}e^{-}e^{+}$, with $N$ and $N^{*}$ the initial an final nuclei) \cite{Murase:2019vdl}. Concretely, at energies below $T_{p} \lesssim 10^4$ GeV, the dominant cooling process is $pp$ interactions with ambient gas. At energies in between $T_{p} \sim 10^4-10^6$ GeV, the Bethe-Heitler production mechanism becomes relevant, and at energies above $\sim 10^6$ GeV, $p\gamma$ interactions are responsible for CR energy losses~\cite{Murase:2022dog}. In TXS 0506+056, CR electrons are mainly cooled by inverse Compton scattering, synchroton radiation, and escape losses \cite{Keivani_2018}. At energies below $T_{e} \lesssim 5$~GeV, escape losses dominate. At energies between $\sim 5$~GeV and $40$~GeV, inverse Compton scattering becomes relevant, and at energies above $\sim 40$~GeV, synchroton radiation becomes the main cooling mechanism of CR electrons. The behaviors of the cooling timescales with the CR energy for both NGC 1068 and TXS 0506+056 are manifest in Fig.~\ref{fig:cooling_timescales}.

In general, the SM cooling processes and timescales for astrophysical sources are model dependent. However, for TXS 0506+056 and NGC 1068, the multimessenger spectral energy distributions are available, enabling us to evaluate these timescales and derive conservative bounds on DM-SM interactions.  

\emph{Cooling timescales induced by inelastic DM-proton and DM-electron interactions.-}
\begin{figure*}
\centering
\includegraphics[width=0.84\textwidth]{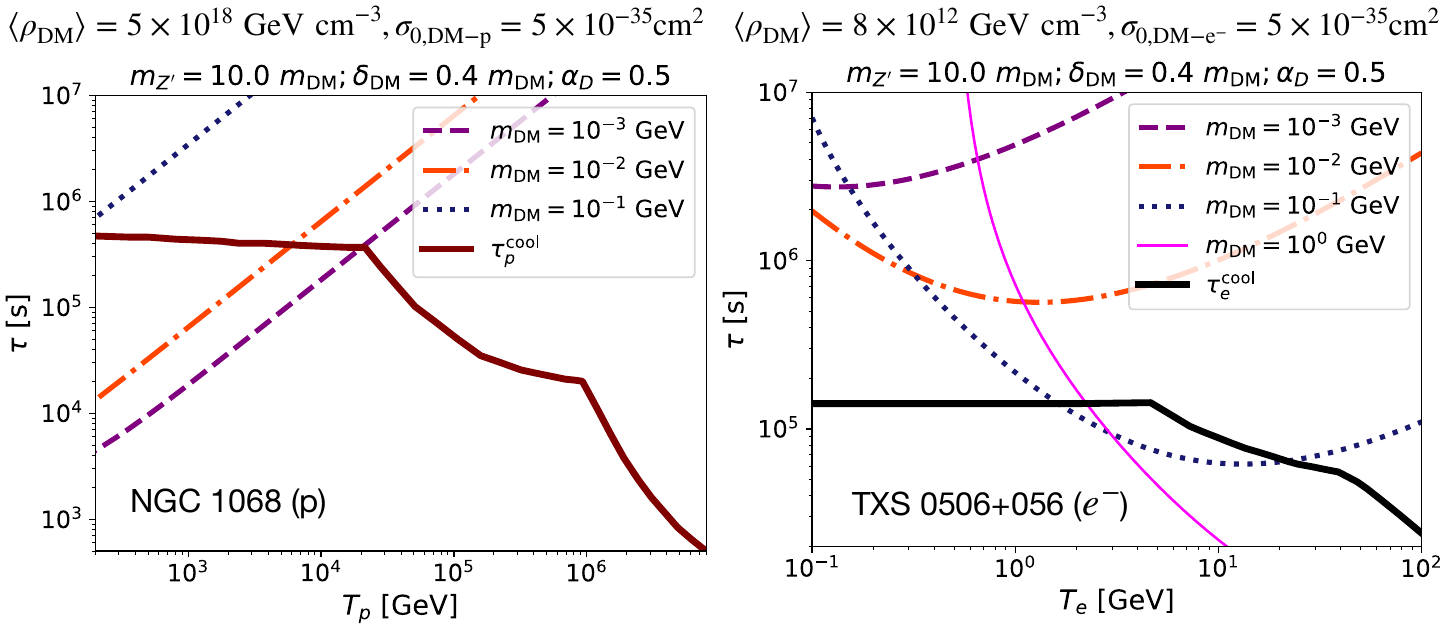}
\centering
\caption{Cooling timescales from CR protons (left) and electrons (right) scattering with inelastic DM, in terms of proton energy $T_p$, compared with those from SM processes~\cite{Murase:2022dog}. We take the average dark matter density obtained from a spike formed in the region of cosmic ray acceleration $\left\langle\rho_{\mathrm{DM}}\right\rangle$ (see the Supplementary Material for details), and fix the non-relativistic scattering cross section to $\sigma_{0 ,\mathrm{DM}-\mathrm{p}}=5 \times 10^{-35} \mathrm{~cm}^2$. We fix the mass splitting among the two dark matter states as $\delta_{\rm DM}=0.4 m_{\rm DM}$, and fix the mediator mass as $m_Z^{\prime}=10m_{\rm DM}$, and sample for different values of the dark matter mass $m_{\rm DM}$. Furthermore, we use a benchmark value of the dark structure constant, $\alpha_{\rm D}=g_{\rm DM}^2/4\pi=0.5$. For the case of proton cooling, we see that the timescales become larger as $m_{\DM}$ increases (i.e., as the number density decreases). The timescales also increase with $T_{p}$ since scattering becomes ineffective at large-momenta transfers. For the case of electron scattering, we see that cooling timescales initially decrease with $T_{e}$, then start to increase once $(2 m_{\DM} T_{e} - \delta_{\rm DM}^2) \gtrsim m^2_{Z'}$.}
\label{fig:cooling_timescales}
\end{figure*}
Neutrino and gamma-ray observations allow us to constrain the CR cooling induced by DM-proton and DM-electron scatterings \cite{Herrera:2023nww}. The cooling time scale due to inelastic DM-SM scatterings is given by 
\begin{equation} \label{eq:InelELoss}
\frac{dE}{d t} = -\frac{\langle \rho_{\rm DM}\rangle}{{m_{\rm DM}}}\,\int_{T^{\rm min}_{\rm DM}}^{T^{\rm max}_{\rm DM}} dT_{\rm DM}\, (T_{\rm DM} + \delta_{\rm DM}) \frac{d\sigma_{\rm DM \, \SM \rightarrow DM^{*} \SM}}{dT_{\rm DM}},
\end{equation}
where $\langle \rho_{\rm DM}\rangle$ is the average DM density in the vicinity of the supermassive back hole at the center of an AGN, where scatterings are more likely to occur. We refer the reader to Supplementary Material Sec.~\ref{sec:details_coolingtimescale} and Sec.~\ref{sec:rhodm} for details on the above equation and $\langle \rho_{\rm DM}\rangle$ respectively \cite{Herrera:2023fpq, Herrera:2023nww, Sigurdsson:2003wu,Merritt:2002vj,Merritt:2006mt,Bertone:2024wbn}. The energy loss rate induced by inelastic DM-proton and DM-electron scatterings can be translated into a cooling timescale of these interactions, which reads
\begin{equation}\label{eq:tau_el}
    \tau^{\rm inel}_{\text{DM}-\SM}=\Bigg[-\frac{1}{E}\bigg(\frac{dE}{dt}\bigg)\Bigg]^{-1}\,.
\end{equation}
\begin{figure*}[t!]
    \centering
	\includegraphics[width=0.49\textwidth]{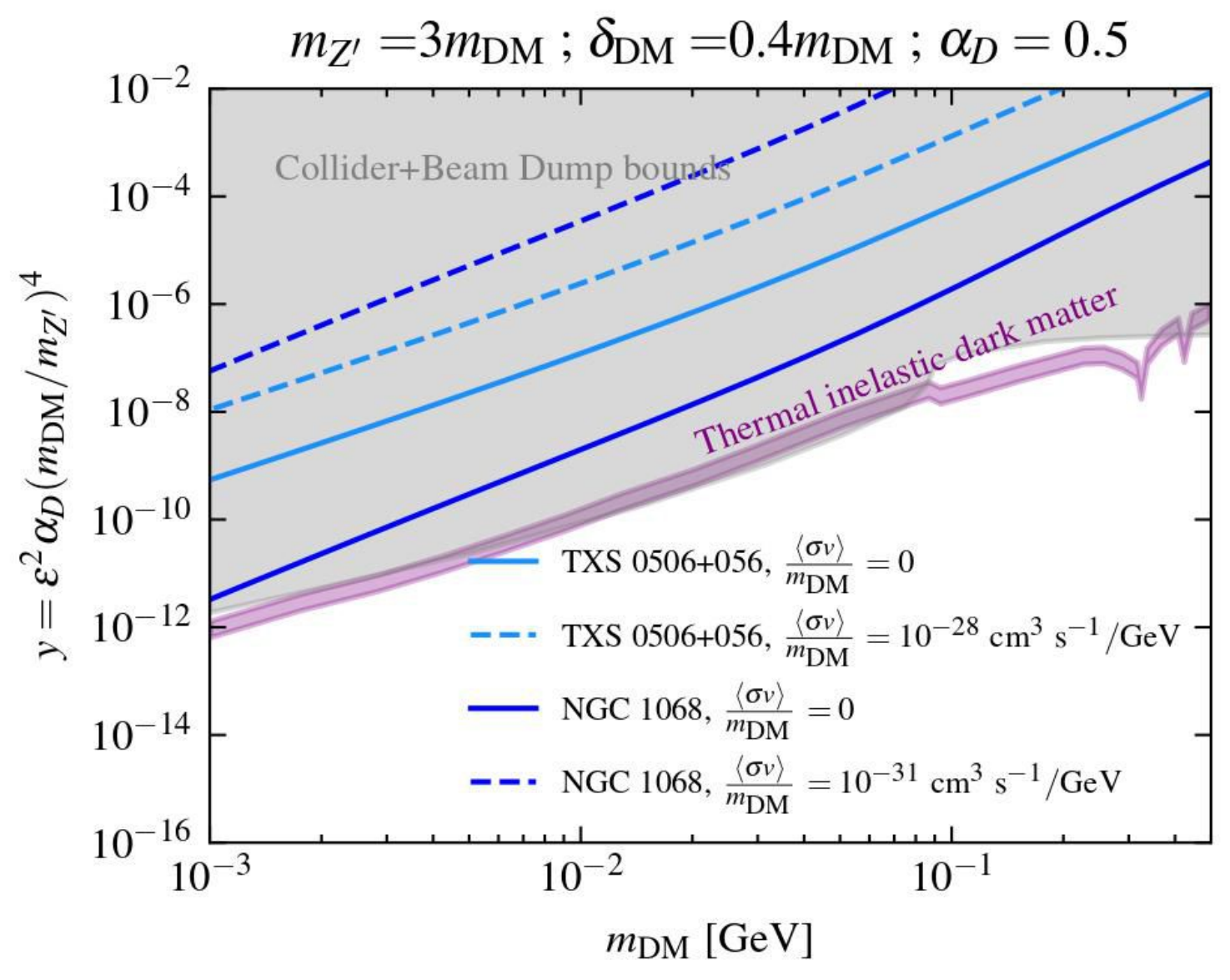}
 	\includegraphics[width=0.49\textwidth]{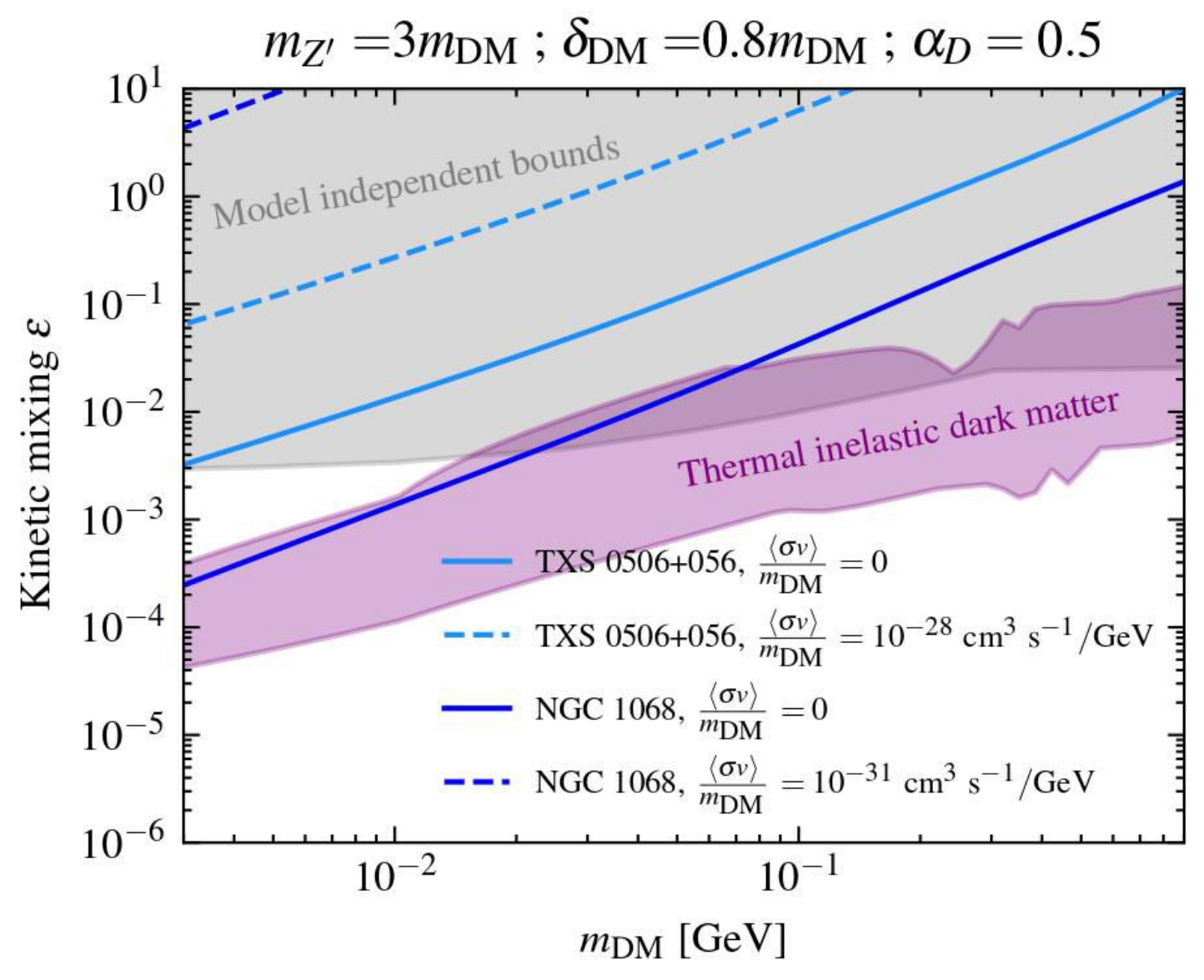}
	\centering
	\caption{\textbf{Left plot:} AGN cooling constraints on the DM-proton and DM-electron interaction strength, parametrized as $y$. Solid lines show constraints when considering the dark matter density at the cosmic ray acceleration regions arising from a spike, while dashed lines correspond to the density expected from a cored profile (see the Supplementary Material for details). For comparison, we show constraints from collider and beam dump experiments, and combination of values able to account for thermal DM, from \cite{Mongillo:2023hbs,Garcia:2024uwf} \textbf{Right plot:} AGN cooling constraints on the kinetic mixing $\epsilon$ vs the mass of the DM $m_{\rm DM}$, for fixed relations $m_{Z'} = 3 m_{\rm DM}$, $\delta_{\rm DM} = 0.8 m_{\rm DM}$ and $\alpha_{\rm DM} = 0.5$. For comparison, we show model-independent constraints from collider and beam-dump experiments, and thermal DM targets from \cite{Garcia:2024uwf}. To derive these constraints, we consider the cooling of cosmic rays in the following energy ranges: 30-100 TeV for NGC 1068, and 0.1-20 PeV for TXS 0506+056 (see main text for details).}
	\label{fig:bounds_light_thermal}
\end{figure*}
In Fig.~\ref{fig:cooling_timescales}, we show the cooling timescales induced DM-proton (electron) interactions~\footnote{For NGC 1068 we do not have significant evidence for electron acceleration and for TXS 0506+056 proton acceleration gives weaker bounds than NGC 1068.} in NGC 1068 (TXS 0506+056), for different values of the DM mass and mass splitting between the two DM states, and fixed benchmark values of the characteristic cross section. For comparison, we show the cooling timescales induced by SM processes in these sources, inferred from multimessenger observations. It can be appreciated that for certain values of the inelastic DM parameters, the cooling timescales can be comparable or shorter than the SM cooling timescales at the relevant energies, which would contradict observations on Earth. Concretely, we can derive an upper limit on the inelastic DM-proton and DM-electron scattering cross section from the requirement
\begin{equation}\label{eq:criteria}
\tau^{\rm inel}_{\text{DM}-\SM}\geq C \, \tau^{\rm cool}_{\SM}\,,
\end{equation}
where $C$ is a model and source dependent factor. We use $C=0.1$ and $C=1$ for NGC 1068 and TXS 0506+056, respectively, to derive constraints in this work. We explain this choice in detail in the Supplementary Material \cite{Das:2024vug, Woo:2002un,Ananna:2020ben,Murase:2022dog,Murase:2018iyl,Keivani_2018}. Briefly, this criterion is consistent with the energetics requirement from multimessenger observations of AGN. For NGC 1068, the CR proton luminosity would be $10^{43}~{\rm erg}~{\rm s}^{-1}\lesssim L_p\lesssim  L_{\rm X-ray}^{\rm tot} \lesssim {\rm a~few}\times 10^{44}~{\rm erg}~{\rm s}^{-1}$~\cite{Murase:2022dog, Das:2024vug}, justifying $C\sim0.1-1$, where $L_X$ is the total (bolometric) luminosity. For TXS 0506+056, the absolute proton luminosity in the single-zone model would violate the Eddington luminosity $L_{\rm Edd}$ \cite{Murase:2018iyl}, so our choice is conservative for protons. This is also reasonable for electrons. The total isotropic equivalent electron luminosity is $L_{e}\sim 8\times{10}^{47}~{\rm erg}~{\rm s}^{-1}$, in which the absolute electron luminosity can be lower than $L_{\rm Edd}$ \cite{Keivani_2018}. For NGC 1068, we consider proton energies 10-300~TeV to place constraints, while for TXS 0506+056 we consider elctron energies 50 GeV - 2 TeV. We note that as $m_{\rm DM}$ increases, the number of DM particles decrease since $\langle \rho_{\rm DM} \rangle$ is fixed leading to higher timescales or lower rates. Thus better constraints can be obtained for lighter DM masses. In fact, we see, for $m_{\rm DM} \sim 10^{-3}$ GeV, the DM induced cooling dominates the SM cooling channels for $T_p < {\rm a\ few} \times 10^4$ GeV. 

We note that, in our model, the upscattered dark matter state may decay into the lightest state plus Standard Model particles, which may re-inject energy into the Standard Model sector affecting non-trivially the electromagnetic spectrum. We considered the decay $\chi_2 \rightarrow \chi_1+\gamma$. While this decay is forbidden at the tree level due to gauge invariance (see our dedicated discussion in the Supplementary Material), it may occur via a dipole operator. Assuming this operator is large enough for the decay to occur on sufficiently small timescales, we estimate the energy transfer into the SM sector to be 1/2 or less for cosmic ray protons, but could be larger for cosmic ray electrons for same combinations of dark matter mass and mass splittings in the dark sector. For the scenario where dark matter cools cosmic ray protons, the energy loss from the scattering will have a direct impact on the energy and flux of resulting high-energy neutrinos, which is a sufficient condition to place our constraints from TXS 0506+056 and NGC 1068 given IceCube data. For dark matter interactions with cosmic ray electrons, however, the modification of the low-energy gamma-ray flux due to decays of the upscattered state may lead to degeneracies with the expected low-energy gamma-ray flux from competing SM processes like electron-positron annihilation, synchroton radiation or Inverse Compton scattering, among others. A calculation of the modified fluxes and their comparison with electromagnetic observations would allow us to refine the ensuing limits derived from cooling arguments in this work.

\begin{figure*}[t!]
    \centering
    \includegraphics[width = 0.47\textwidth]{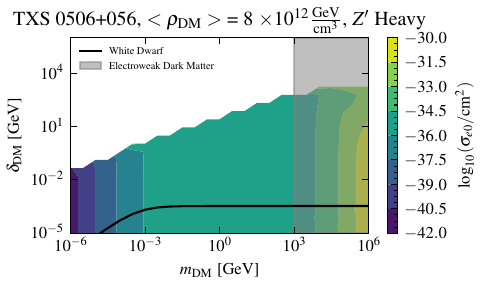}
    \includegraphics[width = 0.47\textwidth]{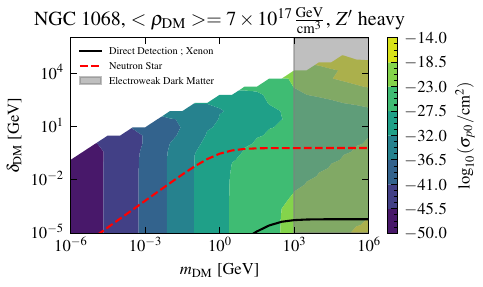}
	\includegraphics[width=0.49\textwidth]{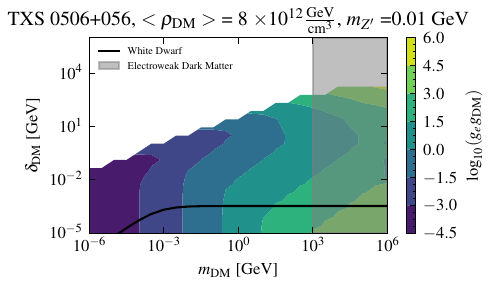}
    \includegraphics[width=0.47\textwidth]{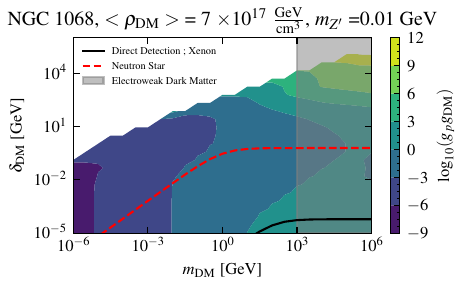}
	\centering
 
    \caption{Upper limits on \textbf{top:} the characteristic scattering cross section given by Eq.~\eqref{eq:CharCrossSec} and \textbf{bottom:} the product of DM and SM couplings to a vector mediator from the cooling of \textbf{left:} electrons in TXS 056-0506 and \textbf{right:} protons in NGC 1068 with a vector mediator. In the top plots, we consider a very heavy mediator (see the supplementary material where we comment on this further), while for the bottom plots we fix the mediator mass. Also included are lines for the maximal mass splitting able to be probed with direct detection and celestial bodies using Eq.~\eqref{eq:delta_req} (direct detection is neglected for electrons, as the electron mass is too small for keV mass splitting with virialized DM). Finally, the region of $m_{\rm DM} > 10^{3}$~GeV is shaded to indicate this region is of special interest for electroweak DM.}
	\label{fig:cooling-contours}
\end{figure*}

\emph{Upper limits on inelastic DM-proton and DM-electron interactions.-} By means of Eq.~\eqref{eq:criteria}, we can derive upper limits on the scattering cross section of inelastic DM off nucleons and electrons. Our model has 5 free parameters: the lighter state DM mass $m_{\rm DM}$, the mass splitting between the DM states $\delta_{\rm DM}$, mediator mass $m_{Z^{\prime}}$ and couplings of the mediator to the SM sector, $g_{\rm SM}$, and to the dark sector $g_{\rm DM}$.

We will derive constraints in this multidimensional parameter space in two distinct and physically motivated regimes. First, we will focus on the parameter space corresponding to a fixed relation between the mediator and DM masses ($m_{\rm Z^{\prime}}=10 m_{\rm DM}$ and $m_{\rm Z^{\prime}}=3 m_{\rm DM}$) and to a fixed relation between the DM mass and the mass splitting ($\delta_{\rm DM}=0.4 m_{\rm DM}$ and $\delta_{\rm DM}=0.8 m_{\rm DM}$). Furthermore, we will fix the dark gauge coupling to a ``natural" value ($\alpha_{\rm D}=0.5$). Such relations between inelastic DM parameters have been discussed previously in the literature, e.g., Refs.~\cite{Izaguirre_2016, Berlin:2018bsc,Mongillo:2023hbs,gonzález2021cosmology, Garcia:2024uwf}. It has been shown that for light (MeV-scale) DM, such relations predict a thermal relic that can be within reach of collider and beam dump experiments. Here we will demonstrate that this region of parameter space can also be probed by CR cooling in AGN.

In Fig.~\ref{fig:bounds_light_thermal} we show constraints in the previously discussed parameter space, from CR proton cooling in NGC 1068 (dark blue) and CR electron cooling in TXS 0506+056 (light blue), for asymmetric or weakly self-annihilating DM (solid) and for a sizable DM self-annihilation cross section ($\langle\sigma v\rangle/m_{\rm DM}=10^{-28} \mathrm{~cm}^3 \mathrm{~s}^{-1} / \mathrm{GeV}$ for TXS 0506+056, $\langle\sigma v/m_{\mathrm{DM}}=10^{-31} \mathrm{~cm}^3 \mathrm{~s}^{-1} / \mathrm{GeV}$ for NGC 1068), which depletes the distribution to a core in these sources. In the left plot, we constrain the quantity $y=\epsilon^2 \alpha_D (m_{\rm DM}/m_{Z^{\prime}})^4$ in which $\epsilon = g_{\SM}/e$, to allow for comparison with existing literature. We show a band of thermal inelastic light DM. The upper end of the band has been derived in various works, e.g., Refs.~\cite{Berlin:2018bsc,Mongillo:2023hbs,gonzález2021cosmology, Garcia:2024uwf}, while the range of values extending to the lowest end was derived in Ref.~\cite{Garcia:2024uwf}. These correspond to different plausible values of the dark left-right coupling asymmetry ($\sim y_L-y_R$) or Majorana mass asymmetry ($\sim m_L-m_R$), which are common parameters in concrete models of inelastic DM. This parameter quantifies the asymmetry in the Lagrangian terms $\mathcal{L}_\chi \supset-\frac{1}{2} m_L \bar{\chi}_L^c \chi_L-\frac{1}{2} m_R \bar{\chi}_R^c \chi_R$ or $\mathcal{L}_\chi \supset-\sqrt{2} y_L S \bar{\chi}_L^c \chi_L-\sqrt{2} y_R S \bar{\chi}_R^c \chi_R$, where $S$ is a singlet scalar and $\chi_{L,R}$ are the left and right DM field components, respectively (see the Supplementary Material for details) \cite{Garcia:2024uwf}. Concretely, denoting the asymmetry as $\delta_y \equiv (y_R-y_L)/y_L=(m_R-m_L)/m_L$, we use values in the range from $\delta_y=0-1000$.

Furthermore, we show in grey a combination of collider and beam dump experiment constraints in these models. For the chosen combination of inelastic DM parameters, our cooling constraints can be stronger than collider constraints for masses below $m_{\rm DM} \sim 1 $~MeV, probing thermal values.

In the right panel of Fig.~\ref{fig:bounds_light_thermal}, we show constraints derived for different values of the mediator and DM mass ratio and mass splitting. In this case, it can be appreciated that our cooling constraints can be stronger than the (model independent) collider and beam dump experimental constraints for masses below $m_{\rm DM} \lesssim 20$~MeV, and allow us to probe thermal inelastic DM for masses below $m_{\rm DM} \lesssim 70$~MeV.

Notably, our limits on $y$ and $\epsilon$ become stronger at light DM masses. This is because: first, the number density of DM particles in the AGN increases for light DM masses, which increases the probability of interactions. Second, the scattering cross section of inelastic DM off electrons and protons in these environments increases at low DM masses, due to the inverse dependence with the reduced mass of the DM-proton and DM-electron systems, confer Eq.~\eqref{eq:Vec_Diff_Sigma}. The enhanced cross section induces shorter cooling timescales at low DM masses, thus stronger bounds on the interaction strength.

The parameter space discussed previously, although predictive, is narrow and may misrepresent the actual relations between the DM mass and the mediator mass, and between the DM mass and the size of the mass splitting. Therefore, in the following we present more general constraints in the parameter space spanned by the DM mass, the mass splitting, and the interaction strength of the DM with the SM sector. We will derive constraints in the limit where the mediator mass of the interaction is much heavier than the momentum transfer of the scattering process in the AGN, and we will present our constraints on the non-relativistic scattering cross section as defined in Eq. (\ref{eq:CharCrossSec}).

Moreover, we present analogous constraints in the limit of a finite mediator mass ($m_{Z^{ \prime}}=$10 MeV), and show contours for limits on the product of the gauge couplings instead of the non-relativistic cross section. It should be noticed that in some regions of the displayed parameter space, particularly at high DM masses, the cooling constraints are very weak, \textit{i.e} it may be difficult to interpret them physically since the cross sections and couplings probed are nonperturbative. A detailed discussion on this regard can be found in the Supplementary Material.

Our results are summarized in Fig.~\ref{fig:cooling-contours}. We show contour constraints on the parameter space of DM mass, mass splitting and scattering cross section, for TXS 0506+056 (left panels) and NGC 1068 (right panels).  We see that cross section bounds become weaker as the DM mass increases (and thus the number density decreases). The bounds become stronger as $\delta_{\rm DM}$ increases because $q^2$ decreases at larger $\delta_{\rm DM}$, giving less suppression to the differential cross section as seen in Eq.~\eqref{eq:Vec_Diff_Sigma}. This effect is even more pronounced for protons, even in the heavy mediator limit, as the form-factor decreases with larger $q^{2}$. We know that cooling is only possible when $\mathbf{s} > (m_{\SM} + m_{\DM} + \delta_{\rm DM})^2$, which is why bounds become weaker when $m_{\DM} \lesssim (\delta_{\rm DM}^2 + 2 m_{\rm SM} \delta_{\rm DM})/(2 T_{\SM,\min} - 2 \delta)$ and completely go away when $m_{\DM} \lesssim (\delta_{\rm DM}^2 + 2 m_{\SM} \delta_{\rm DM})/(2 T_{\SM,\max} - 2 \delta_{\rm DM})$, where $T_{\SM,\min} (T_{\SM, \max})$ is the minimum (maximum) electron or proton kinetic energy considered for cooling.

Inelastic DM has also been explored in the context of direct detection experiments and interactions with compact celestial objects like white dwarfs and neutron stars. In these cases, we may define the kinematic requirement for scattering to be
\begin{equation}
m_{\DM}^2 + m_{\SM}^2 + 2 \frac{m_{\DM} m_{\SM}}{\sqrt{1-v_{rel}^2}} > (m_{\SM} + m_{\DM} + \delta_{\rm DM})^2
\end{equation}
where $v_{rel}$ is the velocity of the DM particle in the rest frame of the SM particle. A detailed analysis of these methods would consider the full velocity distributions of DM and SM particles. For now, we will simply take characteristic relative velocities of $v_{rel} = 10^{-3}$ for direct detection and $v_{rel} = 0.8$ for neutron stars and white dwarfs. This can be translated into a requirement on the mass-splitting of
\begin{equation}
    \delta_{\rm DM} \leq  - m_{\SM} - m_{\DM} 
    + \bigg( m_{\rm SM}^2 + m_{\DM}^2 + \frac{2 m_{\SM} m_{\DM}}{\sqrt{1 - v_{rel}^2}} \bigg)^{1/2}.
    \label{eq:delta_req}
\end{equation}
For comparison purposes, the maximum mass splitting achieved by direct detection experiments and capture in neutron stars and white dwarfs is confronted with our results in Fig.~\ref{fig:cooling-contours}. It can be clearly appreciated that for all DM masses, multimessenger observations of AGN allow us to reach larger mass splittings than complementary probes.

\emph{Discussions and Implications.-}
We have proposed a novel phenomenological probe of inelastic DM, relying on the cooling of CRs in AGN through DM-proton and DM-electron up-scatterings into an excited DM state. DM in the vicinity of NGC 1068 and TXS 0506+056 may scatter off CR protons and electrons, producing a heavier DM in the final state, and cooling the CRs. Since CRs are responsible for the emission of high-energy neutrinos and gamma-rays from these sources, observable on Earth, their cooling timescales would not be dominated by beyond the SM interactions, otherwise such high-energy particle emission would be significantly depleted. The luminosity in protons and electrons in these sources is lower and upper bounded, which allows us to quantify the uncertainties in the inferred cooling timescales within the SM, which are about one order of magnitude in NGC 1068 and TXS 0506+056.

A remarkable result of our work is the demonstration that CR cooling in AGN allows us to constrain unprecedently large mass splittings of inelastic DM particles, above the TeV scale. The cross sections constrained at those mass splittings, however, are larger than those expected in electroweakly interacting DM models. Such mass splittings are orders of magnitude larger than those probed by complementary astrophysical and laboratory searches, as highlighted in Fig.~\ref{fig:schematic}. Furthermore, for thermal light inelastic DM, our cooling constraints can be comparable or even overcome collider and beam dump bounds for DM masses below $m_{\DM} \leq 70$ MeV, probing thermal values.

The fact that DM direct detection experiments, gamma-ray, and neutrino telescopes have not found conclusive evidence for weakly interacting DM may indicate that the portal of the DM to the SM sector is more strongly suppressed than initially expected. The suppression may not occur at the level of the interaction probability, but rather be purely kinematical. If the DM interacts only (or predominantly) inelastically with the SM sector, such process will only occur when the energy transfer of the scattering process exceeds the mass difference between the two DM states. If the mass splitting is sufficiently large, above a few GeV, the prospects for detection with traditional direct detection experiments or astrophysical probes such as neutron stars are not very promising. We have demonstrated that in those scenarios, CR accelerators like AGNs may still allow us to probe those models. A new frontier of inelastic DM with high mass splittings, above the TeV scale, has been opened.

\emph{Acknowledgements.-} 
We are grateful to Felix Kahlhoefer, Thomas Schwetz and Giovani Dalla Valle Garcia for discussions on inelastic DM models. A.G., G.H., and I.M.S. are supported by the U.S. Department of Energy under the award number DE-SC0020262.  M. M. and K. M. are supported by NSF Grant No. AST- 2108466. A.G. is partially supported by the World Premier International Research Center Initiative (WPI), MEXT, Japan. A.G. is grateful to QUP for hospitality during his visit. G.H. and M.M. are grateful to CERN for hospitality during their visit and partial support during the last stages of this work. M.M. also acknowledges support from the Institute for Gravitation and the Cosmos (IGC) Postdoctoral Fellowship. M. M. thanks the Galileo Galilei Institute for Theoretical Physics for the hospitality and the INFN for partial support during the completion of this work. The work of K.M. is supported by the NSF Grant Nos.~AST-2108467, and AST-2308021, and KAKENHI Nos.~20H01901 and 20H05852. 
\bibliographystyle{apsrev4-1}
\bibliography{References.bib}
\clearpage
\newpage
\maketitle
\onecolumngrid
\begin{center}
	\textbf{\Large Supplementary Material}
	 \bigskip\\
		\textbf{\large Cosmic-Ray Cooling in Active Galactic Nuclei as a New Probe of Inelastic Dark Matter}
		 \medskip\\
   {R. Andrew Gustafson, Gonzalo Herrera, Mainak Mukhopadhyay, Kohta Murase, Ian M. Shoemaker}
\end{center}

\renewcommand{\thesection}{S\arabic{section}}
\renewcommand{\theequation}{S\arabic{equation}}
\renewcommand{\thefigure}{S\arabic{figure}}
\renewcommand{\thetable}{S\arabic{table}}
\renewcommand{\thepage}{S\arabic{page}}
\setcounter{equation}{0}
\setcounter{figure}{0}
\setcounter{table}{0}
\setcounter{page}{1}
\setcounter{secnumdepth}{4}

\section{Simplified inelastic dark matter model}
\label{sec:example_model_inelasticDM}
Inelastic dark matter can arise if the dark matter is a fermion with a small Majorana mass in addition to a Dirac mass term, and vector interactions with SM fermions $f$. Given a dark matter fermion $\psi=\left(\eta, \xi \right)$, we can consider the Lagrangian \cite{Tucker-Smith:2001myb}

\begin{equation}
\mathcal{L} \supset g_{\rm DM}\bar{\psi} \gamma_\mu \psi g_{\rm SM}\bar{f} \gamma^\mu  f + m\bar{\psi}\psi+\frac{\delta}{4}(\eta \eta+\bar{\eta} \bar{\eta}).
\end{equation}
Thus assuming $m \gg \delta$, the Majorana fermion mass eigenstates then become
\begin{equation}
\begin{array}{ll}
\chi_1 \simeq \frac{i}{\sqrt{2}}(\eta-\xi) & m_{\chi_1}=m-\delta/2 \\
\chi_2 \simeq \frac{1}{\sqrt{2}}(\eta+\xi) & m_{\rm \chi_2}=m+\delta/2
\end{array}
\end{equation}
and the two dark matter mass states are splitted by $\delta$ in mass. The vector current then leads to the generic case where diagonal interactions among $\chi_1$ and $\chi_2$ are suppressed with respect to the off-diagonal ones by a factor $\delta/4m \ll 1$
\begin{equation}
\bar{\psi} \gamma_\mu \psi \simeq i\left(\bar{\chi}_1 \bar{\sigma}_\mu \chi_2-\bar{\chi}_2 \bar{\sigma}_\mu \chi_1\right)+\frac{\delta}{4 m}\left(\bar{\chi}_2 \bar{\sigma}_\mu \chi_2-\bar{\chi}_1 \bar{\sigma}_\mu \chi_1\right).
\end{equation}
This is one realization of inelastic dark matter. In our phenomenological analysis, we only need to postulate two dark matter states split in mass by $\delta_{\rm DM}$ that undergo off-diagonal interactions mediated by a vector mediator. Such a scenario is present not only in the model here described, but also in several other realizations of inelastic DM, see \textit{e.g} \cite{Tucker-Smith:2001myb, Garcia:2024uwf}

However, in Fig. 3 we show the region of parameter space of light and inelastic dark matter that can reproduce the observed relic abundance in the Universe in a concrete model. We briefly describe this model here, following \cite{Garcia:2024uwf, Kahlhoefer:2015bea, Duerr:2016tmh}. It consists in two dark matter fermions ($\chi_L$, $\chi_R$) with opposite chirality, singlets in the SM and charged under a dark $U(1)^{\prime}$ symmetry, and a scalar field (also singlet under the SM) which breaks the dark gauge symmetry, generating a mass for both the dark matter and the dark vector boson $A^{\prime}$ (In the main text we identify this with the $Z^{\prime}$ vector mediator. The Lagrangian terms involving the dark matter $\chi$, the new mediator $A^{\prime}$ and the scalar $S$ are given by
\begin{align}
\mathcal{L}_{\rm tot}=\mathcal{L}_\chi+\mathcal{L}_{A^{\prime}}+\mathcal{L}_S
\end{align}
with
\begin{align}
\mathcal{L}_\chi  =i \bar{\chi}_L \not D \chi_L+i \bar{\chi}_R \not D \chi_R-m_D^* \bar{\chi}_L \chi_R-\sqrt{2} y_L S \bar{\chi}_L^c \chi_L-\sqrt{2} y_R S \bar{\chi}_R^c \chi_R+\text { h.c. }
\end{align}
\begin{align}
\mathcal{L}_{A^{\prime}}  =-\frac{1}{4} A^{\prime \mu \nu} A_{\mu \nu}^{\prime}-\frac{1}{2} \frac{\epsilon}{\cos \theta} B^{\mu \nu} A_{\mu \nu}^{\prime}
\end{align}
\begin{align}
\mathcal{L}_S  =\left(D^\mu S\right)^*\left(D_\mu S\right)+\mu^2|S|^2-\lambda|S|^4-\lambda_{h}|S|^2|H|^2
\end{align}
where $H$ denotes the SM Higgs field, $B$ refers to the SM hypercharge gauge boson, and the covariant derivative is $i D_\mu S=i \partial_\mu S-q_S g_\chi A_\mu^{\prime} S$. After spontaneous symmetry breaking, the dark vector boson gets a mass $m_{A^{\prime}}=2\sqrt{2}g_{\chi} \langle S\rangle$, and the dark matter fermionic states get masses from the following terms
\begin{equation}
\mathcal{L}_\chi \supset-m_D \bar{\chi}_L \chi_R-\frac{1}{2} m_L \bar{\chi}_L^c \chi_L-\frac{1}{2} m_R \bar{\chi}_R^c \chi_R+\text { h.c. }
\end{equation}
where $m_D$ is the Dirac mass, and the Majorana masses are $m_{L / R}=$ $2 \sqrt{2} y_{L / R} \langle S\rangle$. One can then perform a rotation of the fermion fields, finding two physical Majorana dark matter states, which we will denote by $\chi_1$ and $\chi_2$, with masses
\begin{equation}
m_{\chi_{1,2}}=\sqrt{m_D^2+2 \langle S\rangle^2\left(y_R-y_L\right)^2} \mp 2\langle S\rangle\left(y_L+y_R\right) .
\end{equation}
The same rotation can be applied to the interaction term of the Lagrangian, responsible for setting the dark matter abundance in the Universe. Those interactions can be factorized in terms of the dark-left right Majorana mass or Yukawa asymmetry
\begin{equation}
\delta_y \equiv \frac{y_R-y_L}{y_L}=\frac{m_R-m_L}{m_L}.
\end{equation}
For the thermal bands shown in Figure 3, we take the range $\delta_y=0-1000$, in analogy with \cite{Garcia:2024uwf}.

\section{Forbidden decay $\chi_2 \rightarrow \chi_1 + \gamma$ from the Ward-Takahashi identity}
We analyze the Ward-Takahashi identity for a decay process involving the transition of the heavier dark matter state into the lightest and photon. The Ward-Takahashi identity states that for a conserved current, the matrix element must satisfy
\begin{equation}
    k_\mu \mathcal{M}^\mu = 0 ,\, \, \, \mathcal{M}(k)=\epsilon_\mu(k) \mathcal{M}^\mu(k)
\end{equation}
where $\epsilon_\mu(k)$ is the polarization vector of the photon. The amplitude for the process is given by
\begin{equation}
    \mathcal{M}^\mu = e \, \bar{u}_{\chi_1}(p_2) (-i g \gamma^\mu) u_{\chi_2}(p_1).
\end{equation}
Contracting with the photon momentum $k^\mu$
\begin{equation}
    k_\mu \mathcal{M}^\mu = (-i e g) \bar{u}_{\chi_1}(p_2) k_\mu \gamma^\mu u_{\chi_2}(p_1).
\end{equation}
Using the relation $k^\mu = p_1^\mu - p_2^\mu$, we rewrite
\begin{equation}
    k_\mu \mathcal{M}^\mu = (-i e g) \bar{u}_{\chi_1}(p_2) (\slashed{p}_1 - \slashed{p}_2) u_{\chi_2}(p_1).
\end{equation}
Applying the Dirac equation for the spinors
\begin{equation}
    \slashed{p}_1 u_{\chi_2}(p_1) = m_{\chi_2} u_{\chi_2}(p_1), \quad \bar{u}_{\chi_1}(p_2) \slashed{p}_2 = m_{\chi_1} \bar{u}_{\chi_1}(p_2),
\end{equation}
we obtain
\begin{equation}
    k_\mu \mathcal{M}^\mu = (-i e g) \bar{u}_{\chi_1}(p_2) (m_{\chi_2} - m_{\chi_1}) u_{\chi_2}(p_1).
\end{equation}
Thus, $k_\mu \mathcal{M}^\mu = 0$ is only satisfied if $m_{\chi_2} = m_{\chi_1}$, showing that this tree-level decay channel is not allowed.

\section{Inelastic DM-SM cooling timescale: details for Eq.~(\ref{eq:InelELoss})}
\label{sec:details_coolingtimescale}
In this section we provide details on Eq.~\eqref{eq:InelELoss}. The cooling time scale due to inelastic DM-SM scatterings is given by 
\begin{equation}
\left(\frac{dE}{d t}\right) = -\frac{\langle \rho_{\rm DM}\rangle}{{m_{\rm DM}}}\,\int_{T^{\rm min}_{\rm DM}}^{T^{\rm max}_{\rm DM}} dT_{\rm DM}\, (T_{\rm DM} + \delta_{\rm DM}) \frac{d\sigma_{\rm DM \, \SM \rightarrow DM^{*} \SM}}{dT_{\rm DM}}.
\end{equation}
Here, $\langle \rho_{\rm DM}\rangle$ is the average DM density in the vicinity of the supermassive back hole at the center of an AGN, where scatterings are more likely to occur. These values were calculated under a variety of scenarios in \cite{Herrera:2023fpq}, we refer the reader to \ref{sec:rhodm} for details on this calculation for NGC 1068 and TXS 0506+056. The upper limits of integration denote the minimum and maximum kinetic energy that a DM particle can carry after the collision. To find these energy bounds, let us think of center of mass (COM) before (after) the collision $p_{i}$($p_{f}$) which we can relate to the Mandelstam variable $\mathbf{s} = m_{\SM}^2 + m_{\DM}^2 + 2 (m_{\SM} + T_{\SM})m_{\DM}$ via

\begin{equation} \label{eq:COMpi}
    \mathbf{s} = m_{\SM}^2 + m_{\rm DM}^2 + 2 \bigg( \sqrt{m_{\SM}^2 + p_{i}^2} \sqrt{m_{\rm DM}^2 + p_{i}^2} + p_{i}^2 \bigg),
\end{equation}
before the collision and

\begin{equation} 
\label{eq:COMpf}
\mathbf{s} = m_{\SM}^2 + (m_{\rm DM}+\delta_{\rm DM})^2 + 2 \bigg( \sqrt{m_{\SM}^2 + p_{f}^2} \sqrt{(m_{\rm DM} + \delta_{\rm DM})^2 + p_{f}^2} + p_{f}^2 \bigg)\,,
\end{equation}
afterwards. These are solved by $p_{i}^2 = \frac{1}{4\mathbf{s}}\lambda(m_{\SM}^2, m_{\rm DM}^2, \mathbf{s})$ and $p_{f}^2 = \frac{1}{4 \mathbf{s}}\lambda(m_{\SM}^2, (m_{\rm DM}+\delta_{\rm DM})^2, \mathbf{s})$, where

\begin{equation} \label{eq:KallenFunc}
    \lambda(a,b,c) = a^2 + b^2 + c^2 - 2ab - 2ac - 2bc.
\end{equation}

We can now relate these to the Mandelstam variable \textbf{t} for the cases of maximum and minimum momentum transfer

\begin{equation} 
\label{eq:MandelstamT}
\mathbf{t_{+/-}} = m_{\rm DM}^2 + (m_{\rm DM} + \delta_{\rm DM})^2 - 2 \bigg( \sqrt{m_{\rm DM}^2 + p_{i}^2} \sqrt{(m_{\rm DM} + \delta_{\rm DM})^2 + p_{f}^2} \pm p_{i} p_{f} \bigg)\,,
\end{equation}
which is related to the kinetic energy of the DM in the lab frame (which follows naturally from the definition of \textbf{t} and accounting for the fact that the DM starts at rest) as
\begin{equation}
T_{\rm DM}^{\max /\min} = \frac{\delta_{\rm DM}^2 - \mathbf{t_{+/-}}}{2 m_{\rm DM}}\,.
\end{equation}

Turning now to the cross section, we will consider a vector mediator $Z^{\prime}$. To simplify the cross section formula, the characteristic non-relativistic cross section is defined as
\begin{equation} 
    \sigma_{0} = \frac{g_{\SM}^2  g_{\mathrm{DM}}^2 \mu^2_{\DM-\SM}}{\pi m^{4}_{Z^{\prime}}}\,.
    \label{eq:CharCrossSec}
\end{equation}

The differential cross section for DM scattering with a SM point-particle is
\begin{equation}
    \frac{d \sigma_{pp}}{dT_{\DM}} = \sigma_{0} \frac{m_{Z^{\prime}}^4}{(m_{Z^{\prime}}^2 + q^2)^2} \frac{m_{\DM} \bigg[ \big(\mathbf{s} - (m_{\DM}^2 + m_{\SM}^2 + \delta_{\rm DM} m_{\DM}) \big)^2 + m_{\DM} T_{\DM} (q^2 - 2\mathbf{s}) \bigg] }{2 \mu_{\DM-\SM}^{2} \lambda(m_{\DM}^2,m_{\SM}^2,\mathbf{s})}.
    \label{eq:Vec_Diff_Sigma}
\end{equation}

For protons, we only consider elastic scattering, so we must also include a form-factor,

\begin{equation}
    F_{p}(q^2) = \frac{1}{(1 + q^2/\Lambda^2)^2},
\end{equation}
where $q^{2} = 2 m_{\DM} T_{\DM} - \delta_{\rm DM}^2$, $\Lambda \approx$ 770 MeV, and the cross section for elastic proton scattering goes as $d \sigma_{p}/dT_{\DM} = F^2_{p}(q^2) d\sigma_{pp}/dT_{\DM}$.

It is useful to relate the non-relativistic scattering cross section with the microphysics of a concrete model. For now, we will let $g_{\SM} = \epsilon e$ where $e$ is the fundamental electric charge. (This is often considered for the case of a ``dark photon" below the electroweak scale, but note that for dark photon masses above the electroweak scale, the coupling is proportional to the hypercharge). We will also use the parameter $\alpha_{\rm DM} = g_{\rm DM}^2/4\pi$.

\section{DM distribution in NGC 1068 and TXS 0506+056}
\label{sec:rhodm}
\begin{figure}
\centering
\includegraphics[width=0.49\textwidth]{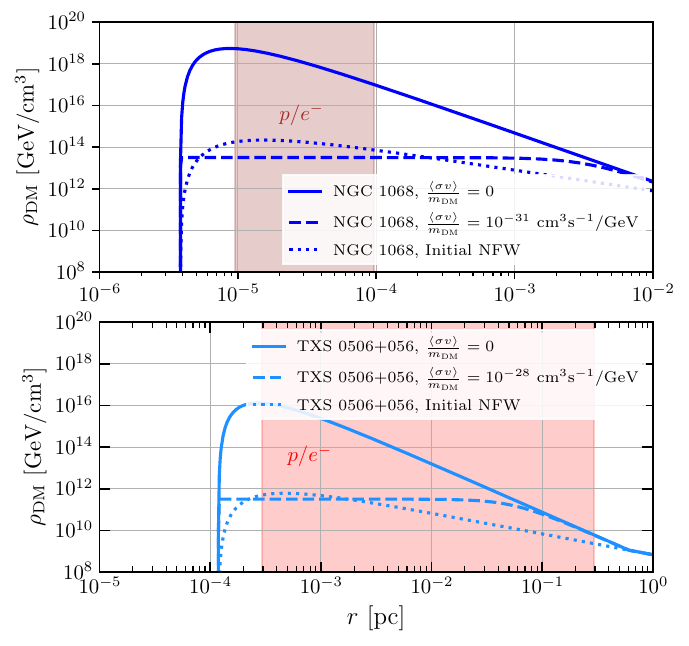}
\caption{DM profiles corresponding to NGC 1068 (\emph{top panel}) and TXS 0506+056 (\emph{bottom panel}). The profiles are shown for both when DM annihilation cross-sections leads to a sizable flattening of the DM density (dashed lines) and when DM annihilation cross-sections are small or zero and the spike is preserved (solid lines). For comparison, we show the initial NFW profile from which the DM spike is formed. The typical emission regions are also shown and shaded with maroon for NGC 1068 and red for TXS 0506+056. Profiles for different values of the initial NFW-like profile index $\gamma$ and a discussion their associated uncertainties can be found in \cite{Herrera:2023nww}.
}
\label{fig:dmprof}
\end{figure}
\begin{table*}
		\begin{center}
			\begin{tabular}{l|ccccccc}
				\hline
				& $R_{\rm em}$ &  $M_{\rm BH}$ & $t_{\rm BH}$ & $r_0$ & $\langle\sigma v \rangle$/$m_{\rm DM}$ & $\langle \rho_{\rm DM} \rangle$  \\
				\hline
				\toprule
				 NGC 1068 (I) & 30 $R_S$  & (1-2) $\times 10^{7}M_{\odot}$ &  $10^{10}$ yr & 10 kpc & 0 & $5\times10^{18}$ GeV/cm$^{3}$   \\
				 NGC 1068 (II) & 30 $R_S$ & (1-2) $\times 10^{7}M_{\odot}$ &  $10^{10}$ yr & 10 kpc & $10^{-31}$cm$^{3}$s$^{-1}$/GeV & $4\times10^{13}$ GeV/cm$^{3}$      \\
				 TXS 0506+056 (I) & $10^4 R_S$ &  (3-10) $\times 10^{8}M_{\odot}$ &  $10^9$ yr & 10 kpc & 0 & $8\times10^{12}$ GeV/cm$^{3}$  \\
				 TXS 0506+056 (II) & $10^4 R_S$ & (3-10) $\times 10^{8}M_{\odot}$ &  $10^9$ yr & 10 kpc &$10^{-28}$cm$^{3}$s$^{-1}$/GeV & $4\times10^{11}$ GeV/cm$^{3}$   \\
                \hline
			\end{tabular}
		\end{center}
		\caption{Relevant parameters considered in this work for NGC 1068 and TXS 0506+056, for two different sets of assumptions dubbed (I) and (II). Here $R_{\rm em}$ represents the distance of the emission region from the central SMBH in NGC 1068 (TXS 0506+056), $M_{\rm BH}$ shows the SMBH mass and its uncertainty, $t_{\rm BH}$ is the black hole age, $r_0$ is the scale radius of the galaxy, $\langle\sigma v \rangle$/$m_{\rm DM}$ denotes the assumed values of the effective DM self-annihilation cross section, and $\langle \rho_{\rm DM} \rangle$ is the average density of DM particles within $R_{\rm em}$.}
		\label{tab:AGN}
	\end{table*}
In this section we discuss the distribution of DM in the sources that we consider for this work - NGC 1068 and TXS 0506+056. It is evident from Eq.(2) that the relevant quantity for DM-induced cooling timescales is the average DM density $\langle \rho_{\rm DM} \rangle$ in the vicinity of the SMBH that the protons can scatter off. Thus the DM distribution around these sources need to be modeled to then evaluate $\langle \rho_{\rm DM} \rangle$.

We assume that the central SMBH has an adiabatic growth in the DM background. This is reasonable and can be understood by comparing the dynamical timescale ($t_{\rm dyn}$) with the Salpeter timescale  ($t_{\rm sal}$)~\cite{Sigurdsson:2003wu}. The dynamical timescale, $t_{\rm dyn} = G M_{\rm BH}/\sigma_v^3$, where $M_{\rm BH}$ is  the mass of the SMBH and $\sigma_v$ is the velocity dispersion of stars outside the radius of influence. The Salpeter timescale is associated with the BH growth and is given by $t_{\rm sal} = M_{\rm BH}/\dot{M}_{\rm Edd}$, where $\dot{M}_{\rm Edd}$ is the Eddington accretion timescale. For most SMBHs, $t_{\rm dyn} \ll t_{\rm sal}$ justifying the assumption of adiabatic growth.

Let us begin by considering a typical NFW profile with index $\gamma=1$. This initial profile eventually evolves into a spike profile with $\gamma_{\rm sp}$ in its inner regions. The density of the spike can be given as
\begin{equation}
\rho_{\mathrm{sp}}(r)=\rho_R g_\gamma(r)\left(\frac{R_{\mathrm{sp}}}{r}\right)^{\gamma_{\mathrm{sp}}},
\end{equation}
where $r$ is the radial distance from the center of the SMBH, the normalization factor $\rho_R=\rho_0\left(R_{\mathrm{sp}} / r_0\right)^{-\gamma}$, is chosen such that the density profile outside of the spike matches with that of the boundary of the spike. We choose the scale radius $r_0 = 10$ kpc, that is, galactic scales. The size of the spike is given by $R_{\mathrm{sp}}=\alpha_\gamma r_0\left( M_{\mathrm{BH}} /(\rho_0 r_0^3)\right)^{\frac{1}{3-\gamma}}$. The \emph{cuspiness} of the spike is given by $\gamma_{\mathrm{sp}}=\frac{9-2 \gamma}{4-\gamma}$. The parameter $\alpha_\gamma \simeq 0.293 \gamma^{4 / 9}$ for $\gamma \ll 1$ and $g_\gamma(r)$ can be approximated for $0<\gamma<2$ by $g_\gamma(r) \approx\left(1-\frac{4 R_{\mathrm{s}}}{r}\right)$. It is trivial to note that this density profile vanishes at 4 $R_S$, where $R_S = 2GM_{\rm BH}/c^2$ is the Scwarzschild radius of the SMBH. 

So far we did not take into account DM self-annihilations. In some instances, the DM annihilation cross-section can be sufficiently large to affect the DM spike. The DM density profile then flattens and saturates to $\rho_{\text {sat }}=m_{\mathrm{DM}} /\left(\langle\sigma v\rangle t_{\mathrm{BH}}\right)$, where $\langle\sigma v\rangle$ is the velocity-averaged DM self-annihilation cross section, and $t_{\mathrm{BH}}$ is the age of the SMBH. The spike profile  extends to a maximal radius $R_{\mathrm{sp}}$, beyond which the DM distribution follows the NFW profile. Putting it altogether we have the DM density profile modeled as
\begin{equation}
\rho(r)= \begin{cases}0 & r \leq 4 R_{\mathrm{S}} \\ \frac{\rho_{\mathrm{sp}}(r) \rho_{\mathrm{sat}}}{\rho_{\mathrm{sp}}(r)+\rho_{\mathrm{sat}}} & 4 R_{\mathrm{S}} \leq r \leq R_{\mathrm{sp}}, \\ \rho_0\left(\frac{r}{r_0}\right)^{-\gamma}\left(1+\frac{r}{r_0}\right)^{-(3-\gamma)} & r \geq R_{\mathrm{sp}} .\end{cases}
\end{equation}
The parameters we chose for modeling the DM density for the two sources are shown in Table~\ref{tab:AGN} and the resulting DM density profiles as a function of the radial distance are shown in Fig.~\ref{fig:dmprof}. We also show the neutrino and gamma-ray emission regions with a shaded band. We note that the accelerated cosmic rays (CRs) would encounter a large portion of the DM profile which is evident from the overlap between the shaded band and the DM density curves. The constraints will however be stronger in the presence of a spike due to larger values of $\langle \rho_{\rm DM} \rangle$. This is manifest in Fig.~\ref{fig:bounds_light_thermal} of our work.

Besides self-annihilations, the dark matter spike profile steepness may be relaxed by other processes. Mergers of galaxies can form supermassive black hole binaries which transfer energy to the the dark matter spike, reducing the steepness of the density profile. However, it is not generically the case that the spike is completely erased back into a core, see \textit{e.g} \cite{Merritt:2002vj}. It has been further discussed that the dark matter spike may be heated by gravitational interactions with baryons, reducing the spike index to a value as low as $\gamma_{\rm sp}=1.5$, see \textit{e.g} \cite{Merritt:2006mt}. Moreover, in a recent work~\cite{Bertone:2024wbn} the evolution of dark matter overdensities was discussed in the context of modified NFW profile through the formation of a supermassive star followed by its collapse to a black hole and finally the growth of the black hole to its final mass. Of course, such dynamic evolution models may modify our results which are beyond the scope of the current work. The reduction on the dark matter density profile in these scenarios is typically smaller than the reduction which would be induced by sizable self-annihilations. Therefore, our upper limits derived in this work for a depleted core can be considered robust against such uncertainties. Furthermore, even if adiabatic growth does not occur and the spike is not even formed, an extrapolation to lower scales of an NFW profile, as expected from N-body simulations of $\Lambda$CDM, leads to a larger density than a spike depleted to a core by self-annihilations. Lastly, the supernovae explosions affecting the spike are typically at distances of $\mathcal{O}(1)\ \rm kpc$ from the galactic center, while the region of particle acceleration are constrained to be at the sub-parsec level alleviating this issue.

\section{Uncertainties in the CR luminosity}

In this section we address the uncertainties associated with the CR proton and electron luminosities at the source. This helps in quantifying and justifying the choice of $C$ in Eq.~\eqref{eq:criteria}. The maximum CR luminosity associated with AGN can be estimated based on various theoretical and observational arguments that we discuss below.


The Eddington luminosity ($L_{\rm edd}$) can be computed by balancing the outward radiation pressure and the inward gravitational force. We thus have in hydrostatic equilibrium
\begin{equation}
L_{\rm edd} = \frac{4 \pi G c m_{\mathrm{p}}}{\sigma_T} M_{\rm BH} = (1.3 \times 10^{45})\ {\rm erg\ s}^{-1} \left( \frac{M_{\rm BH}}{10^7\ M_\odot} \right)\,,
\end{equation}
where $M_{\rm BH}$ is the mass of the black hole, $\sigma_T$ is the Thomson cross-section. The Eddington luminosity if violated is mostly for very short periods of time for most sources. 
Now, assuming that the accretion power is converted to accelerate the CR protons, we have
\begin{equation}
\label{eq:accpow}
\int L_{\varepsilon_p}d\varepsilon_p = \eta_{\rm CR} \dot{M}c^2\,,
\end{equation}
where $\eta_{\rm CR}$ is the acceleration efficiency, $\dot{M}$ is the mass accretion rate, and $\epsilon_p$ is the cosmic ray proton energy in the comoving frame. We can estimate the value of $\eta_{\rm CR}$ in the following way. The gravitational energy released in the process of accretion is shared by the non-thermal and thermal emissions, bulk motion, CR protons, and various other emission processes. Assuming that all CR protons are utilized for radiation, from the virial theorem ($2T + U = 0$, where $T$ is the kinetic energy and $U$ is the potential energy) we have for a given distance from the central black hole $R$,
\begin{equation}
2 \eta_{\rm CR} \dot{M}c^2 = \frac{G M_{\rm BH} \dot{M}}{R}\,.
\end{equation}
Massaging the above equation we find
\begin{equation}
\eta_{\rm CR} = 0.025 \left( \frac{M_{\rm BH}}{10^7\ M_\odot} \right) \left( \frac{R}{10\ R_S} \right)^{-1}\,,
\end{equation}
where the Schwarzschild radius is given by $R_S$. Thus for NGC 1068 (TXS 0506+056), $\eta_{\rm CR} \sim 0.01\ (0.0002)$ assuming the emission radius $R_{\rm em} = 30\ R_S\  (10^4\ R_S)$ and $M_{\rm BH} = 10^7\ M_\odot (10^8\ M_\odot)$.
Using the well-observed X-ray luminosity, for NGC 1068 we have the total coronal luminosity above $2$ keV as $L^{\rm tot}_{\rm X-ray} = 7.84 \times 10^{43}\ {\rm erg\ s}^{-1}$~\cite{Das:2024vug}. Besides the X-rays from the hot corona there is an additional component from the optically thick and geometrically thin disk such that the bolometric luminosity is $L_{\rm bol} = 4.82 \times 10^{44}\ {\rm erg\ s}^{-1}$~\cite{Woo:2002un}. Thus the Eddington parameter becomes $\lambda_{\rm Edd} = L_{\rm bol}/L_{\rm Edd} \sim 0.4 - 0.5$. Assuming an accretion efficiency of $\eta_{\rm rad} = 0.1$~\cite{Ananna:2020ben}, we have $\dot{M}c^2 = L_{\rm bol}/\eta_{\rm rad} = 4.82 \times 10^{45}\ {\rm erg\ s}^{-1}$. Thus from Eq.~\eqref{eq:accpow}, we have for the CR protons, $L_{\rm CR} \sim 5 \times 10^{43}\ {\rm erg\ s}^{-1}$. This combination of theoretical constraints and observational signatures help us reasonably claim that for CR protons in NGC 1068 we have $10^{43}\ {\rm erg\ s}^{-1} \leq L_p \leq L_{\rm X-ray}^{\rm tot} \leq {\rm a\ few} \times 10^{44}\ {\rm erg\ s}^{-1}$ (see below Eq.~\ref{eq:criteria}). Furthermore, based on the observed differential neutrino luminosity around 1 TeV ($\varepsilon_\nu L_{\varepsilon_\nu} \sim 3 \times 10^{42}\ {\rm erg\ s}^{-1}$) the inferred $L^{\nu-\rm obs}_{\rm CR} \sim 8 \times 10^{42}\ {\rm erg\ s}^{-1}$~\cite{Murase:2022dog} which is $\sim 10$ times smaller than the $L_{\rm CR}$ we estimate above.

Even after this to be conservative, we still leave room for roughly \emph{an order of magnitude} in the CR luminosity uncertainty associated with NGC 1068, by choosing $C = 0.1$. This means that our limits are derived by imposing that the dark matter-induced cooling timescales shall not overcome the observationally inferred timescales by more than one order of magnitude.

Similarly for TXS 0506+056, following~\cite{Murase:2018iyl} we argue that the absolute proton luminosity in the single-zone model would violate $L_{\rm Edd}$, so our choice is conservative. However, for TXS 0506+056 the limits from cosmic ray electrons are more stringent. In this case, we have from~\cite{Keivani_2018} $L_e \sim 8 \times 10^{47}\ {\rm erg\ s}^{-1} \sim 20\ L_{\rm Edd}$, thus choosing $C = 1$, that is, timescales comparable to the Standard Model timescales, is justified.

\section{Cooling timescales}
In this section we focus on some relevant details associated with the cooling timescales. In Fig.~\ref{fig:cooling_timescales} we showed a typical case for the case of a finite mediator mass. Below we consider two case studies concerning a very heavy mediator and an alternate finite mediator.
\subsection{Case study: heavy mediator}
The results of the cooling timescales considering a very heavy mediator $m_{Z^\prime}$ is shown in the \emph{upper} panels of Fig.~\ref{fig:cooling_timescales_supp}. We note that for NGC 1068, the timescales are lower by a few orders of magnitude as compared to the case of a finite mediator mass $m_{Z^\prime} = 10 m_{\rm DM}$ as illustrated in Fig.~\ref{fig:cooling_timescales} of the main draft. Thus the constraints become stronger when a infinitely heavy mediator is considered.

For TXS 0506+056 on the other hand, the effects of the mediator mass are insignificant for lower values of electron kinetic energies as compared to the finite mediator case. The decrease of the cooling timescales with increase in $T_e$ is also consistent to the finite mediator case. However, recall that for the finite mediator case, the timescales increase for $2m_{\rm DM}T_e - \delta_{\rm DM}^2 \gtrsim m_{Z^\prime}^2$, which is not applicable here since $m_{Z^\prime} \gg m_{\rm DM}$. As a result, the timescales decrease with time without having a turn over. 
\subsection{Case study: alternate finite mediator mass}
\begin{figure}
\centering
\includegraphics[width=0.9\textwidth]{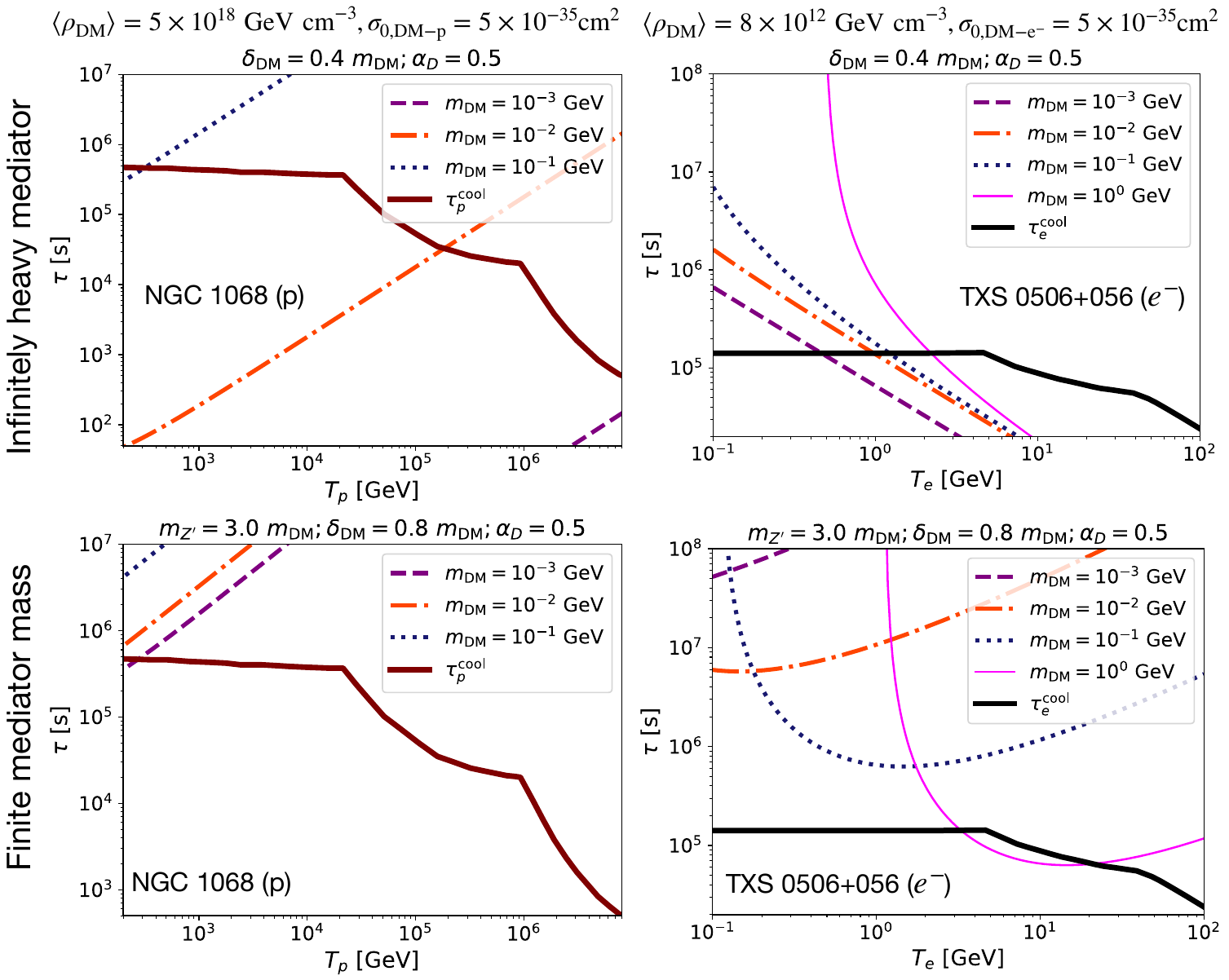}
\caption{Cooling timescales from scattering with inelastic DM, compared with cooling from SM processes: \emph{upper panels:} considering a very heavy mediator $m_{Z^\prime}$ and \emph{lower panels:} considering a finite mass mediator $m_{Z^\prime} = 3\ m_{\rm DM}$.
}
\label{fig:cooling_timescales_supp}
\end{figure}
We illustrated a representative case for finite mass mediator where we set $m_{Z^\prime} = 10 m_{\rm DM}$ and the mass splitting $\delta_{\rm DM} = 0.4 m_{\rm DM}$. In this subsection, we present an alternate parameter set for the finite mediator case to highlight the differences. We choose $m_{Z^\prime} = 3\ m_{\rm DM}$ and $\delta_{\rm DM} = 0.8\ m_{\rm DM}$. Our results are shown in the \emph{lower} panels of Fig.~\ref{fig:cooling_timescales_supp}. We notice that these parameter choices would lead to weaker constraints as compared to what is presented in Fig.~\ref{fig:cooling_timescales}.

\subsection{Scanning Parameter Space}
\begin{figure*}
    \centering
    \includegraphics[width = 0.49 \textwidth]{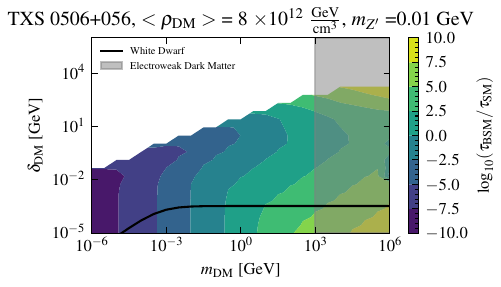}
    \includegraphics[width = 0.49 \textwidth]{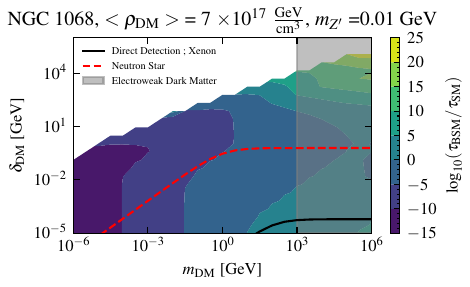}
    \caption{Minimum ratio of the cooling timescale from DM scattering to the SM cooling timescale. Here we have fixed $g_{\SM} = g_{\DM} =1$. \label{fig:Cooling_Contour}}
\end{figure*}

In order to obtain our bounds, for each value of $m_{\DM}$ and $\delta_{\rm DM}$, we must find the minimum ratio of the BSM cooling timescale to the SM cooling timescale for the energies considered. We have included examples of such ratios in Fig. \ref{fig:Cooling_Contour}, where we have made a guess of the couplings to be $(g_{\SM} g_{\DM})_{\rm guess} = 1$. We provide our limits on the couplings by stating $(g_{\SM} g_{\DM})_{\rm exc} = (g_{\SM} g_{\DM})_{\rm guess} * \sqrt{\tau_{\rm BSM}/\tau_{\SM}}$.

\section{Viability of Heavy Mediators \label{sec:Heavy}}
\begin{figure}
\includegraphics[width=0.475\textwidth]{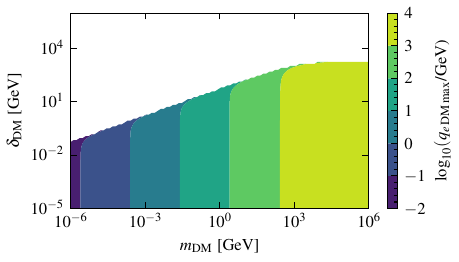}
\includegraphics[width=0.475\textwidth]{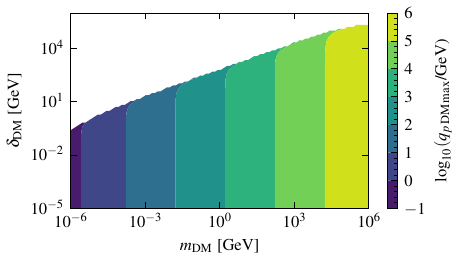}
\caption{\textbf{Left:} Maximum momentum transfer for electrons scattering with inelastic DM. \textbf{Right:} Maximum momentum transfer for protons scattering with inelastic DM. \label{fig:max_momentum_transfer}}
\end{figure}

For simplicity, we have included plots where we claim that the mediator mass is ``heavy". Physically, this corresponds to $m_{Z^{\prime}} \gg q = \sqrt{-\mathbf{t}}$, which depends on both the initial and final energies. To demonstrate the most extreme cases, we consider $q_{\max} =\sqrt{-\mathbf{t_{+}}}$ calculated for the maximum incoming kinetic energy considered ($T_{e,\max} = 2$ TeV, $T_{p, \max} = 300$ TeV). We show the results in Fig. \ref{fig:max_momentum_transfer}.

If we want $m_{Z^{\prime}} > q_{\max}$ and our model to be perturbative (i.e. $g_{\SM}, g_{\DM} < \sqrt{4 \pi}$), this sets a maximum cross section of

\begin{equation}
    \sigma_{\max} = \frac{(4\pi)^2 \mu_{\DM-\SM}^2}{\pi q_{\max}^4}.
\end{equation}

\begin{figure}
\includegraphics[width=0.475\linewidth]{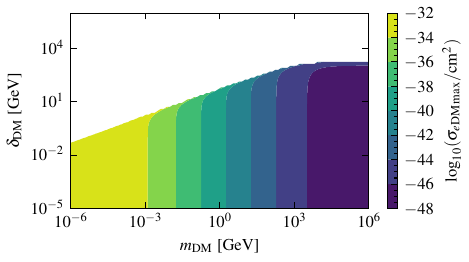}
\includegraphics[width=0.475\textwidth]{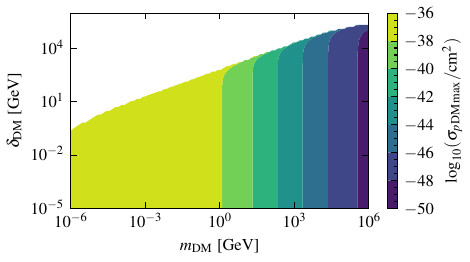}
\caption{\textbf{Left:} Maximum cross section for electron-DM scattering satisfying the ``infinite" mass mediator and perturbative requirements. \textbf{Right:} Same as left, but for proton-$\DM$ scattering \label{fig:max_pert_cross_sec}}
\end{figure}

This cross section is given in Fig.~\ref{fig:max_pert_cross_sec}. Comparing with the bounds given in Fig.~\ref{fig:cooling-contours}, we see that for low DM masses, our bounds are below the perturbativity limits. However, for larger DM masses, in the heavy mediator limit, we can only exclude couplings larger than $\sqrt{4 \pi}$.

\end{document}